\documentclass{article}
\usepackage{amsmath, amssymb, amsfonts}
\title{Hyperbolic vacuum decay} 
\author{Hristu Culetu, \\Ovidius University, Dept.of Physics and Electronics, \\B-dul Mamaia 124, 900527 Constanta, Romania, \\e-mail : hculetu@yahoo.com}

\begin{document}
\numberwithin{equation}{section}
\pagenumbering{arabic}
\maketitle
\newcommand{\fv}{\boldsymbol{f}}
\newcommand{\tv}{\boldsymbol{t}}
\newcommand{\gv}{\boldsymbol{g}}
\newcommand{\OV}{\boldsymbol{O}}
\newcommand{\wv}{\boldsymbol{w}}
\newcommand{\WV}{\boldsymbol{W}}
\newcommand{\NV}{\boldsymbol{N}}
\newcommand{\hv}{\boldsymbol{h}}
\newcommand{\yv}{\boldsymbol{y}}
\newcommand{\RE}{\textrm{Re}}
\newcommand{\IM}{\textrm{Im}}
\newcommand{\rot}{\textrm{rot}}
\newcommand{\dv}{\boldsymbol{d}}
\newcommand{\grad}{\textrm{grad}}
\newcommand{\Tr}{\textrm{Tr}}
\newcommand{\ua}{\uparrow}
\newcommand{\da}{\downarrow}
\newcommand{\ct}{\textrm{const}}
\newcommand{\xv}{\boldsymbol{x}}
\newcommand{\mv}{\boldsymbol{m}}
\newcommand{\rv}{\boldsymbol{r}}
\newcommand{\kv}{\boldsymbol{k}}
\newcommand{\VE}{\boldsymbol{V}}
\newcommand{\sv}{\boldsymbol{s}}
\newcommand{\RV}{\boldsymbol{R}}
\newcommand{\pv}{\boldsymbol{p}}
\newcommand{\PV}{\boldsymbol{P}}
\newcommand{\EV}{\boldsymbol{E}}
\newcommand{\DV}{\boldsymbol{D}}
\newcommand{\BV}{\boldsymbol{B}}
\newcommand{\HV}{\boldsymbol{H}}
\newcommand{\MV}{\boldsymbol{M}}
\newcommand{\be}{\begin{equation}}
\newcommand{\ee}{\end{equation}}
\newcommand{\ba}{\begin{eqnarray}}
\newcommand{\ea}{\end{eqnarray}}
\newcommand{\bq}{\begin{eqnarray*}}
\newcommand{\eq}{\end{eqnarray*}}
\newcommand{\pa}{\partial}
\newcommand{\f}{\frac}
\newcommand{\FV}{\boldsymbol{F}}
\newcommand{\ve}{\boldsymbol{v}}
\newcommand{\AV}{\boldsymbol{A}}
\newcommand{\jv}{\boldsymbol{j}}
\newcommand{\LV}{\boldsymbol{L}}
\newcommand{\SV}{\boldsymbol{S}}
\newcommand{\av}{\boldsymbol{a}}
\newcommand{\qv}{\boldsymbol{q}}
\newcommand{\QV}{\boldsymbol{Q}}
\newcommand{\ev}{\boldsymbol{e}}
\newcommand{\uv}{\boldsymbol{u}}
\newcommand{\KV}{\boldsymbol{K}}
\newcommand{\ro}{\boldsymbol{\rho}}
\newcommand{\si}{\boldsymbol{\sigma}}
\newcommand{\thv}{\boldsymbol{\theta}}
\newcommand{\bv}{\boldsymbol{b}}
\newcommand{\JV}{\boldsymbol{J}}
\newcommand{\nv}{\boldsymbol{n}}
\newcommand{\lv}{\boldsymbol{l}}
\newcommand{\om}{\boldsymbol{\omega}}
\newcommand{\Om}{\boldsymbol{\Omega}}
\newcommand{\Piv}{\boldsymbol{\Pi}}
\newcommand{\UV}{\boldsymbol{U}}
\newcommand{\iv}{\boldsymbol{i}}
\newcommand{\nuv}{\boldsymbol{\nu}}
\newcommand{\muv}{\boldsymbol{\mu}}
\newcommand{\lm}{\boldsymbol{\lambda}}
\newcommand{\Lm}{\boldsymbol{\Lambda}}
\newcommand{\opsi}{\overline{\psi}}
\renewcommand{\tan}{\textrm{tg}}
\renewcommand{\cot}{\textrm{ctg}}
\renewcommand{\sinh}{\textrm{sh}}
\renewcommand{\cosh}{\textrm{ch}}
\renewcommand{\tanh}{\textrm{th}}
\renewcommand{\coth}{\textrm{cth}}

\begin{abstract}
The properties of an hyperbolically-expanding  wormhole are studied. Using a particular equation of state for the fluid on the wormhole throat, we reached an equation of motion for the throat that leads to a constant surface energy density $\sigma$. The Lagrangean leading to the above equation of motion contains the ''rest mass'' of the expanding particle as a potential energy. The associated Hamiltonian corresponds to a relativistic free particle of a total Planck energy $E_{P}$. When the wormhole is embedded in de Sitter space, we found that the cosmological constant is of Planck order but hidden at very tiny scales, in accordance with Carlip's recipe.
 \end{abstract}
 
\section{Introduction}
As everything gravitates, we have to keep track of the vacuum energy, the energy of quantum fluctuations of empty space. Its gravitational effects should manifest as a cosmological constant $\Lambda$ \cite{SC1}. The problem is that simple evaluations give a cosmological constant (CC) which is 120 orders of magnitude larger than what observes at the cosmological scale (for example, from the accelerated expansion of the Universe). As Carlip \cite{SC2} (see also \cite{PR}) has noticed, the trouble comes from the mixing of scales. $\Lambda$ is generated at Planck scale but observed at cosmological scale. He proposed a new, simple alternative: our Universe does have a CC of the order of Planck's value, due to the ''spacetime foam'' picture \cite{JW}, which is not, of course, homogeneous, so that the model is not ruled out by observations.

The nontrivial topological structure of the spacetime at the Planck scale enforces the presence of wormholes (WH), i.e. hypersurfaces that connect two asymptoticall-flat spacetimes \cite{MT1, MT2, MV1, MV2, HC}. Redmount and Suen \cite{RS1, RS2} considered a ''Minkowski WH geometry'' as a model for topological fluctuations within the Planck scale spacetime foam. Their simple WH geometry seems to be unstable against grouth to macroscopic size. 

We investigate in this paper a slightly modified version of the dynamic RS wormhole, taking a different equation of state of the boundary fluid and show that its throat evolves hyperbolically. The equation of motion of the expanding ''bubble'' of the time-dependent mass is a hyperbola which macroscopically becomes the Minkowski light cone. We built the gravitational action from which the equation of motion may be derived and found the corresponding Hamiltonian which proves to be a constant (Planck energy).

Throughout the paper geometrical units $G = c = \hbar = 1$ are used, excepting otherwise specified.

\section{Equation of motion of the throat} 
  Although at macroscopic scales the spacetime appears smooth and simply connected, on Planck length scales it fluctuates quantum-mechanically, developing all kinds of topological structures, including WHs. A microscopic WH may be extracted from the foam to give birth to a macroscopic traversable WH. 
	Redmount and Suen \cite{RS1, RS2} see Lorentzian spacetime filled with many microscopic WHs. They found those WHs are quantum-mechanically unstable, like a classical stable black hole which however undergoes quantum Hawking evaporation. 
	RS constructed a spherically-symmetric ''Minkowski wormhole'' by excising a sphere of radius $r = R(t)$ ($t$ - the Minkowski time coordinate) from two copies of the Minkowski space, identifying the two boundary surfaces $r = R(t)$. To obey Einstein's equations a surface stress tensor on the boundary $\Sigma$ was introduced. Outside the boundary both exterior spacetimes are flat. The boundary plays the role of the WH throat and the Einstein equations are equivalent with the Lanczos equations \cite{MV1} 
  \begin{equation}
  -8\pi S^{i}_{~j} = \left[K^{i}_{~j} - \delta^{i}_{~j}K^{l}_{~l}\right]
 \label{2.1}
 \end{equation}    
with $S^{i}_{~j}$ the surface stress tensor (here $i, j = \tau, \theta, \phi$), $K^{l}_{~l}$ - the trace of the extrinsic curvature of the boundary $\Sigma$ and $[...]$ stands for the jump of $K^{i}_{~j}$ when the boundary is crossed; namely, $[K_{ij}] = K_{ij}^{+} - K_{ij}^{-} = 2 K_{ij}^{+}$ in our situation. 

 Let us find now the extrinsic curvature tensor of the surface $F \equiv r - R(t) = 0$. The spacetime metric is Minkowskian and the geometry on $\Sigma$ can be written as
  \begin{equation}
ds^{2}_{\Sigma} = - d\tau^{2} + R^{2}d\Omega^{2}
 \label{2.2}
 \end{equation}   
where $d\tau = \sqrt{1 - \dot{R}^{2}}dt$, $\tau$ is the proper time on $\Sigma$ and $\dot{R} = dR/dt$. The velocity 4-vector is 
  \begin{equation}
  u^{b} = \frac{dx^{b}}{d\tau} = \left(\frac{dt}{d\tau}, \frac{dR}{d\tau}, 0, 0\right)
 \label{2.3}
 \end{equation}   
with $u^{b}u_{b} = -1$. The unit normal to $\Sigma$ may be found from (2.3) and the relations $n^{b}n_{b} = 1$ and $n^{b}u_{b} = 0$. The velocity $u^{b}$ from (2.3) yields
  \begin{equation}
  u^{b} = \left(\frac{1}{\sqrt{1 - \dot{R}^{2}}}, \frac{\dot{R}}{\sqrt{1 - \dot{R}^{2}}}, 0, 0\right)
 \label{2.4}
 \end{equation}   
whence
  \begin{equation}
  n_{b} = \left(-\frac{\dot{R}}{\sqrt{1 - \dot{R}^{2}}}, \frac{1}{\sqrt{1 - \dot{R}^{2}}}, 0, 0\right)
 \label{2.5}
 \end{equation} 
The second fundamental form of $\Sigma$ may be obtained from \cite{LL}
  \begin{equation}
  K_{ij} = \frac{\partial x^{a}}{\partial \xi^{i}}\frac{\partial x^{b}}{\partial \xi^{j}}\nabla_{a}n_{b},
 \label{2.6}
 \end{equation}      
where $\xi^{i} = (\tau, \theta, \phi)$ are the coordinates on $\Sigma$ and the operator ''nabla'' is applied in Minkowski space in four dimensions.
 
Eq. (2.6) gives the following components of the second fundamental form 
  \begin{equation}
  K_{\tau \tau} = -\frac{\ddot{R}}{(1 - \dot{R}^{2})^{3/2}}, ~~~K_{\theta \theta} = \frac{R}{\sqrt{1 - \dot{R}^{2}}} = \frac{K_{\phi \phi}}{sin^{2}\theta},
 \label{2.7}
 \end{equation}   
with the trace
  \begin{equation}
 K \equiv K^{i}_{~i} = \frac{\ddot{R}}{(1 - \dot{R}^{2})^{3/2}} + \frac{2}{R \sqrt{1 - \dot{R}^{2}}}.
 \label{2.8}
 \end{equation}   
Using (2.7) and (2.8), Eqs. (2.1) give us
   \begin{equation}
  S_{\tau \tau} = -\frac{1}{2\pi R\sqrt{1 - \dot{R}^{2}}}, ~~~S_{\theta \theta} = \frac{R}{4\pi \sqrt{1 - \dot{R}^{2}}} + \frac{R^{2}\ddot{R}}{4\pi (1 - \dot{R}^{2})^{3/2}} = \frac{S_{\phi \phi}}{sin^{2}\theta}.
 \label{2.9}
 \end{equation}   
 Supposing that $S_{ij}$ on the throat corresponds to a perfect fluid
   \begin{equation}
   S_{ij} = (p_{s} + \sigma)u_{i}u_{j} + p_{s}h_{ij}
 \label{2.10}
 \end{equation}   
 where $h_{ij} = diag(-1, R^{2}, R^{2}sin^{2}\theta)$ is the metric on the boundary, we have $\sigma = S_{\tau \tau}$ for the surface energy density and $p_{s} = S_{\theta \theta}/R^{2}$ for the surface pressure. 
 
 To find the equation of motion for the throat, we need now an equation of state relating $\sigma$ and $p_{s}$. RS chose $\sigma = -4p_{s}$ as equation of state but in this case the action integral (whence the equation of motion was obtained) has a complicate ''kinetic term''. Our choice for the equation of state is simply $p_{s} = -\sigma$, as for a domain wall \cite{IS} because one seems to be the most appropriate conjecture for a Lorentz-invariant vacuum. That choice leads to the equation of motion 
    \begin{equation}
    R \ddot{R} + \dot{R}^{2} - 1 = 0,
 \label{2.11}
 \end{equation}   
 which has the solution
    \begin{equation}
    R(t) = \sqrt{t^{2} + b^{2}},
 \label{2.12}
 \end{equation}   
using appropriate initial conditions ($R_{min} = b$, which is taken to be the Planck length).

\section{Free particle energy}

 By means of (2.9), the expression (2.12) for $R(t)$ yields
    \begin{equation}
    \sigma = -p_{s} = -\frac{1}{2\pi b} = const.
 \label{3.1}
 \end{equation}  

 From (2.12) we also have 
 \begin{equation}
 1 - \dot{R}^{2} = \frac{b^{2}}{R^{2}}, ~~~\ddot{R} = \frac{b^{2}}{R^{3}}.
 \label{3.2}
 \end{equation}
 Therefore, the component of the acceleration of the throat, normal to $\Sigma$ will be \cite{MV1}
  \begin{equation}
  A_{\bot} \equiv n_{b}A^{b} = -K_{\tau \tau} = \frac{1}{b} = -2\pi \sigma > 0,
 \label{3.3}
 \end{equation}
 where $A^{b}$ is built with $u^{b}$ from (2.4). So we obtained the same evolution of the WH throat as Ipser and Sikivie for their domain wall which in Minkowskian coordinates is not a plane at all but rather an accelerating sphere, expanding with the acceleration $2\pi |\sigma|$.
 
 A remark is in order here. The radial null geodesics (6.4) from \cite{HC} are similar with the equation of motion (2.12) of the dynamic WH throat. Note that the spacetime (2.2) from \cite{HC} is curved and the region $r < b$ is absent from the manifold. We identify the two processes and assume that actually the null particles are carried by the WH throat during their propagation (see also \cite{HC1}). In other words, the throat turns out to play the role of a de Broglie pilot wave, dragging the null particles with it. The observed absence of macroscopic WHs may be due to their very fast expansion ($R(t) \approx t$ for $t >> b$) and their energy is spread out on larger and larger volumes.
 
 The gravitational action corresponding to (2.12), obtained from the integral of the scalar curvature plus the surface term, may be written as
    \begin{equation}
    S = -\int{\frac{b^{2}}{R}\sqrt{1 - \dot{R}^{2}}dt },
 \label{3.4}
 \end{equation}   
whence the Lagrangean is given by 
    \begin{equation}
   L = -\frac{b^{2}}{R} \sqrt{1 - \dot{R}^{2}}                              
 \label{3.5}
 \end{equation}   
(the factor $b^{2}$ is necessary for $L$ to get units of length). The action (3.4) gives a model with features like those of a relativistic free particle. Moreover, when $S$ - which is invariant - is expressed in terms of the proper time, with $0\leq \tau <\infty$, one obtains $\pi \hbar/2 \geq S>0$.
  
	It is worth noting that, when $\dot{R} <<1$, $L$ acquires the form
	    \begin{equation}
   L \approx -\frac{b^{2}}{R} (1 - \frac{\dot{R}^{2}}{2}) = \frac{Mv^{2}}{2} - Mc^{2},
 \label{3.6}
 \end{equation}   
where $v(t) = \dot{R}$ and $M(t) = \frac{b^{2}}{R(t)}$. We observe that the second term from the r.h.s. of (3.6) plays the role of a time dependent ''rest'' (potential) energy of the expanding ''particle''. 
	
	The canonical momentum will be
    \begin{equation}
  p = \frac{\partial \textit{L}}{\partial \dot{R}} = \frac{b^{2}\dot{R}}{R \sqrt{1 - \dot{R}^{2}}} = \frac{Mv}{\sqrt{1 - v^{2}}}
 \label{3.7}
 \end{equation}   
which yields the Hamiltonian
    \begin{equation}
  H = p \dot{R} - L = \frac{b^{2}}{R \sqrt{1 - \dot{R}^{2}}} = \frac{M}{\sqrt{1 - v^{2}}}
 \label{3.8}
 \end{equation}   
To find the direct relation between $p$ and $H$ we get rid of $\dot{R}$ from the last two equations to obtain
  \begin{equation}
  H = \sqrt{p^{2} + \left(\frac{b^{2}}{R}\right)^{2}}.
 \label{3.9}
 \end{equation}  
Inserting all fundamental constants, we again see that $b^{2}/R = \hbar/cR$ plays the role of a mass $M(t)$ of the ''particle'' (expanding WH throat in our case), namely $M(t) = \hbar/cR(t)$. So $R(t) = \hbar/M(t)c$ appears to be the Compton wavelength associated to the mass $M(t)$. For $t >> b$, $R(t) \approx t$ so that $Mc^{2}t = \hbar$, which looks like an uncertainty relation. One could also see from (3.8) that $H$ has the form of a Lorentz-boosted energy. This simple WH geometry seems to represent a spacetime foam structure unstable against growth to macroscopic size \cite{RS1}.
When (2.12) is used in the expression for the expanding throat energy (3.8), we get $H = b = E_{P}$, where $E_{P}$ is the Planck energy. In other words, the total energy of the expanding ''bubble'' remains constant, although the mass $M$ and $p$ are time-dependent. That is possible because, while $M$ decreases with time, $p$ increases and so there is a perfect compensation between them.
  When the action (3.4) is expressed in terms of the proper time and by means of (2.12), one obtains
	    \begin{equation}
    S = -\int{\frac{b}{cosh\frac{\tau}{b}}d\tau } = b^{2} arcsin\left(\frac{1}{cosh\frac{\tau}{b}}\right),
 \label{3.10}
 \end{equation}   
with $0\leq \tau <\infty$. We get that $\pi \hbar /2 >S>0$. In addition, as a function of variable $\tau$ only, $S$ is an invariant.

\section{The cosmological constant}
Let us consider now the previous WH embedded in a de Sitter (dS) spacetime. We are no longer dealing with a ''Minkowski wormhole''. The ''dS wormhole'' is obtained by excising a sphere of radius $r = R(t)$ from two copies of the $dS$ space. Outside the boundary both exterior spaces are $dS$, with the same horizon radius. Due to the horizon, some restrictions will be imposed on $R(\tau)$, where $\tau$ is the proper time on the throat.

The junction conditions for the system de Sitter - Schwarzschild have been studied, among others, by Blau, Guendelman and Guth \cite{BGG}, so that we will use their model and adopt the calculations in our situation. As we stated previously, we have $[K_{ij}] = K_{ij}^{+} - K_{ij}^{-} = 2 K_{ij}^{+}$ in our conditions, because of the symmetry (see, for example, \cite{GLV}). Keeping in mind that we study the motion of the WH throat (a domain wall with $p_{s} = -\sigma >0$), we get from the junction condition for $K_{\theta \theta}$
   \begin{equation}
	2\sqrt{1 - \chi^{2}R^{2} + R'^{2}} = -4\pi \sigma R,
 \label{4.1}
 \end{equation}  
where $R' = dR/d\tau,~\chi^{2} = \Lambda /3$ and $\Lambda > 0$ is the cosmological constant. We notice that (4.1) has $R(\tau) = b~cosh(\tau/b)$ as a solution (with $R(0) = b$ as initial condition) provided
   \begin{equation}
	 b^{2}(\chi^{2} + g^{2}) = 1,
 \label{4.2}
 \end{equation}  
with the acceleration $g = 2\pi |\sigma|$, $\chi < 1/b$ (or $\Lambda <3/b^{2}$) and $g < 1/b$.

The metric on the WH throat may be written now
    \begin{equation}
		ds^{2}_{\Sigma} = - d\tau^{2} + b^{2}cosh^{2}\frac{\tau}{b}d\Omega^{2},
 \label{4.3}
 \end{equation}  
which is the closed dS space in three dimensions. It is worth noting that three parameters are encountered here: $\chi, g$ and $b$. Dealing with WHs expanding from the Planck world, $b$ has been chosen of the order of the Planck length. From (4.2) we get $\chi = (1/b) \sqrt{1 - b^{2} g^{2}}$ whence one sees that $\Lambda$ reaches its Planck value when $g << 1/b$. 

Let us express now the expanding WH throat  radius in terms of the coordinate time $t$ (the equivalent of Eq.2.12). From the static dS metric \cite{BGG} and the equation of the surface $\Sigma,~r = R(\tau)$, one obtains that
    \begin{equation}
		\frac{dt}{d\tau} = \frac{\sqrt{1 - \chi^{2}R^{2} + R'^{2}}}{1 - \chi^{2}R^{2}}
 \label{4.4}
 \end{equation}  
whence
    \begin{equation}
		\frac{dt}{d\tau} = \frac{\sqrt{1 - \chi^{2}b^{2}}~ cosh \frac{\tau}{b}}{1 - \chi^{2}b^{2} cosh^{2} \frac{\tau}{b}}
 \label{4.5}
 \end{equation}  
with $cosh \frac{\tau}{b} < 1/b\chi$. Eq. 4.5 could be easily integrated and gives us
    \begin{equation}
		\frac{(\alpha + \beta tanh\frac{\tau}{2b})~(\beta - \alpha tanh\frac{\tau}{2b})}{(\alpha - \beta tanh\frac{\tau}{2b})~(\beta + \alpha tanh\frac{\tau}{2b})} = e^{2\chi t},
 \label{4.6}
 \end{equation}  
with $tanh (\tau/2b) < \alpha/\beta,~\tau(0) = 0,~\alpha = \sqrt{1 - b\chi}$ and $\beta = \sqrt{1 + b\chi}$. We have further
    \begin{equation}
		\frac{\chi}{g}~sinh~\frac{\tau}{b} = tanh~\chi t,
 \label{4.7}
 \end{equation}  
that yields
    \begin{equation}
		R(t) = b \sqrt{1 + \frac{g^{2}}{\chi^{2}} tanh^{2}~\chi t}. 
 \label{4.8}
 \end{equation}  
As a consistency check, we take above $\chi = 0$ (the Minkowski case). Having $0/0$ in (4.8), we ought to consider the limit when $\chi \rightarrow 0$ and get
     \begin{equation}
		R(t) = b \sqrt{1 + g^{2} t^{2}}. 
 \label{4.9}
 \end{equation}  
 But $\chi = 0$ gives $g = 1/b$ and so the equation of motion (2.12) is recovered, as expected.

We know that the condition $R < 1/\chi$ must be obeyed. In addition, $\chi < 1/b$ and, therefore, one obtains $R_{max} = (1 + b\chi )/2\chi$. In the asymptotically flat situation ($\chi = 0$), we get $R_{max} \rightarrow \infty$, as it should be according to (2.12), when $t \rightarrow \infty$.

Let us check now the energy constraint equation\cite{MGLV}, obtained from the 3+1 decomposition of Einstein's equations, when the matter contribution is overlooked w.r.t. $\Lambda$
 \begin{equation}
	\begin{split}
	^{3}R - K^{2} + K^{i}_{~j}K^{j}_{~i} - 2\Lambda = 0,
	\end{split}
 \label{4.10}
 \end{equation}  
where $^{3}R$ stands for the curvature scalar of $\Sigma$ with $g_{ij}$ from (2.2). Thanks to the equation $R(\tau ) = b~cosh~\frac{\tau}{b}$, one obtains
     \begin{equation}
 K^{\tau}_{~\tau} = \frac{R'' - \chi^{2}R}{\sqrt{1 - \chi^{2}R^{2} + R'^{2}}} = g,~~~ K^{\theta}_{~\theta} =  \frac{\sqrt{1 - \chi^{2}R^{2} + R'^{2}}}{R} = g,
 \label{4.11}
 \end{equation}  
whence $K = 3g$ and $K^{i}_{~j}K^{j}_{~i} = 3g^{2}$. From (2.2) we have $^{3}R = 6/b^{2}$. Keeping in mind that $\Lambda = 3/b^{2} - 3g^{2}$, one finds that Eq.4.10 is observed.
As far as the shear tensor 
  \begin{equation}
	\sigma^{i}_{~j} = K^{i}_{~j} - \frac{1}{3}\delta^{i}_{~j} K
 \label{4.12}
 \end{equation}  
is concerned, using the components of the extrinsic curvature tensor one finds that $\sigma^{i}_{~j}$ is vanishing. In contrast, the expansion scalar of the fluid on $\Sigma$ is constant, with $\Theta = K = 3g$. 
 
We have seen above that $\Lambda$ takes Planck order values. To measure it, we have to perform experiments at Planck scale, using an apparatus of the same order of magnitude; or, the duration of the measurements to be alike. Similar difficulties arise when one intends to measure the difference between Eq. (2.12) and the standard equation of the light cone $R(t) = t$ in Minkowski space: $b$ is too small and the hyperbola is very close to its asymptotes. 

 \section{Conclusions}
 We investigated in this paper a particular dynamic Lorenzian WH. Using a different equation of state compared to Redmount and Suen, we found that the dynamic WH expands hyperbolically in a fashion similar with the Coleman and de Luccia bubble \cite{CL} or Ipser and Sikivie domain wall, i.e. a Lorentz-invariant expansion. In addition, the Hamiltonian of the system equals the Planck energy $b$ and corresponds to a relativistic free particle of a time-dependent mass $M = \hbar/cR(t)$ and a time-dependent momentum $p = bt/R(t)$, in spite of the fact that its energy is constant. When $|\sigma|$ is small w.r.t. $1/2\pi b$, a Planck-valued $\Lambda$ emerges, in the case our WH is embedded in dS spacetime, which is in accordance with Carlip's conjecture that the CC is huge but hidden at very tiny scales.

\end{document}